# Rapid Electron Backscatter Diffraction Mapping: Painting by Numbers

Vivian S. Tong[1,2], Alexander J. Knowles[1], David Dye[1], T. Ben Britton[1]*
1. Department of Materials, Imperial College London, Exhibition Road, London, United Kingdom, SW7 2AZ
2. Now at the National Physical Laboratory, Teddington, TW11 0LW
*b.britton@imperial.ac.uk

## Abstract

Microstructure characterisation has been greatly enhanced through the use of electron backscatter diffraction (EBSD), where rich maps are generated through analysis of the crystal phase and orientation in the scanning electron microscope (SEM). Conventional EBSD analysis involves raster scanning of the electron beam and the serial analysis of each diffraction pattern in turn. For grain shape, crystallographic texture, and microstructure analysis this can be inefficient. In this work, we present Rapid EBSD, a data fusion approach combining forescatter electron (FSE) imaging with static sparse sampling of EBSD patterns. We segment the FSE image into regions of similar colour (i.e. phase and crystal orientation) and then collect representative EBSD data for each segmented region. This enables microstructural assessment to be performed at the spatial resolution of the (fast) FSE imaging whilst including orientation and phase information from EBSD analysis of representative points. We demonstrate the Rapid EBSD technique on samples of a cobalt based superalloy and a strained dual phase titanium alloy, comparing the results with conventional analysis. Rapid EBSD is advantageous for assessing grain size distributions in time-limited experiments.

## Highlights

1. We segment colour forescatter electron images into distinct regions of common contrast.
2. For each region, obtain one EBSD pattern to measure phase and crystal orientation.
3. Our static sparse sampling method is reconstructed, providing 100x speed up for microstructure mapping.

## Graphical Abstract

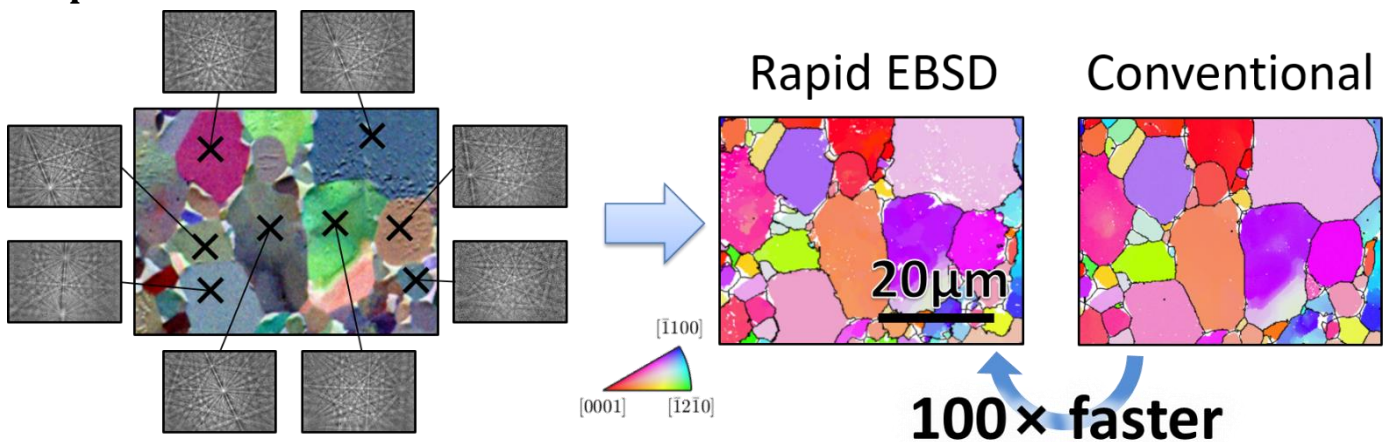

# 1    Introduction

Electron backscatter diffraction (EBSD) is a scanning electron microscope (SEM)-based technique used to characterise the microstructure of crystalline materials. It has a high spatial resolution on the order of 20-100 nm and can also map large areas on the order of several millimetres for a beam scanning set up. In a typical setup, the electron beam is rastered across a highly tilted sample in a regular grid. Typically, diffraction patterns are collected at between 40 – 3000 patterns per second, and indexed to build up a dataset containing orientation and phase information at each data point. Representative microstructure maps are reconstructed from analysis of the EBSD patterns to characterise microstructure. For more information on the technique the reader is directed to reviews by Wilkinson and Britton [1], and Dingley [2].

In cases where we know a significant amount of information about the sample and our proposed microstructural analysis scheme, the regular grid sampling is an oversampling of microstructural information. Regular grid sampling is sub-optimal if different spatial resolutions are required for crystallographic and morphological features in the microstructure. For example, to link grain boundary morphology to (mis)orientation in microtexture analysis, high spatial resolution is needed to identify the boundary trace, but only representative grain orientations are needed. In such cases, the EBSD analysis strategy can be significantly accelerated through sparse sampling of the EBSD field of view, and then reconstructing microstructure information from this representative data.

Rich microstructure maps can be obtained through high contrast observation of orientation and phase contrast using back- and fore- scatter electrons (using electron channelling contrast imaging, ECCI e.g. in Reference [3]). These images can be optimised to provide qualitative orientation and morphology data if the imaging conditions are tuned to produce high contrast between crystal orientations in the imaging field of view. Electron channelling electron imaging has several forms, notably: backscatter electron imaging at low sample tilt using a backscatter electron detector, EBSD pattern-based imaging [4] in EBSD acquisition geometry, and backscatter and forescatter electron imaging at high sample tilt using imaging diodes [3] mounted at the top and bottom of the EBSD detector, respectively (Figure 1).

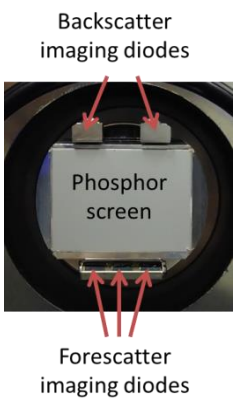

**Figure 1: EBSD detector showing positions of backscatter forescatter imaging diodes relative to the phosphor screen.**

Typically these ECCI images can be captured very rapidly (at a few µs per sampling point, as compared to a few ms per EBSD pattern capture) which is advantageous for high spatial resolution and rapid microstructural analysis. However, quantitative assessment of these images is challenging as variations in contrast are rarely quantitatively described, due the complicated physics involved in generating contrast. However, ECCI has proven popular in microstructural characterisation where ECCI micrographs are used to generate average values of grain size using intercept or planimetric methods [5].

The redundancy of the EBSD regular grid sampling and the lack of quantitative information within ECCI micrographs represents a significant opportunity for data fusion. This has motivated us to generate a sparse sampling method which uses the ECCI image to target representative points for EBSD analysis, and subsequently reconstruct microstructure maps.

Sparse sampling can be performed using either (1) static or (2) dynamic methods.

(1) Static sparse sampling methods follow a sampling scheme determined before starting the experiment, and include random (or pseudo-random) sampling, low-discrepancy sampling, and Lissajous trajectory scanning. Random sampling can be refined by using low-discrepancy sampling [6] to ensure even distribution of sampling points. Lissajous trajectory sampling [7] uses curved scan trajectories which allow quick sampling of the entire field of view, and result in progressively increasing spatial resolution with increasing scan time. Full images can be reconstructed from the acquired sparse data using compressive sensing techniques, such as in-painting of missing pixels and total variation-based de-noising [8–10]. These methods often require around 10% sampled fraction to reconstruct the image, and reproduction of higher order features require higher sampling fractions [10].

(2) Dynamic sparse sampling methods determine the location of the next sampling point 'on the fly' by analysing already collected data. One method for dynamic sparse sampling is to use feature tracking of an initial coarse scan to threshold 'interesting' areas for sampling at higher spatial resolution. This has been applied successfully to confocal laser scanning microscopy [11] in images where interesting features produce more intense fluorescence signal. These approaches have not be used widely in EBSD, where we have only found dynamic sampling used in the machine learning algorithm 'supervised learning approach for dynamic sampling', or SLADS [12]. The SLADS algorithm is pre-trained using similar microstructures, so as to choose sampling points which minimise the estimated image distortion in the reconstructed image. When applied to EBSD data, this method samples densely near grain boundaries and sparsely at grain interiors; successful reconstruction can be achieved with a sampling fraction of 6 % (after pre-training). Variants of this algorithm have been demonstrated to successfully reconstruct electron dispersive X-ray spectroscopy data and SEM images [12,13].

In this work, we describe the Rapid EBSD method, which uses static sparse sampling informed by qualitative forscatter ECCI imaging. This method builds upon high speed orientation mapping using 'colour orientation contrast imaging', originally developed by Day and Quested [14]. Based upon our prior work [3], we optimise the FSE detector positions to optimise contrast for segmentation. We subsequently collect representative EBSD data from each segmented region centre. We reconstruct the final microstructure maps offline.

This static sparse sampling strategy provides the spatial resolution of the ECCI image and the orientation information of the EBSD technique, which adds value to cases where orientation and morphology need to be correlated. For example where grain boundary morphologies are of interest, the grain size distribution is multimodal, measuring grain size resolved texture, or grain boundary misorientation distributions, e.g. in small island grains surrounded by large grains, a microstructure typical in abnormal grain growth, and seen in silicon steels [15–17], aluminium welds [18–21], and zirconium alloys [22,23].

We note that single point per grain sampling does not add extra value where orientation and morphology can be measured separately (e.g. independent macrotexture and average grain intercept length measurements). However instrument and user time costs can be render this approach attractive, especially when considering the robustness of grain size assessment using a threshold grain boundary misorientation angle and ambiguity in contrast in the ECCI images. For texture analysis, the Rapid EBSD technique is attractive when spatial correlation of texture may be of interest (e.g. in microstructure gradients, grain size or phase based texture correlations).

## 2    Algorithm

In this section we present the fundamental aspects of the RapidEBSD algorithm which consists of two major aspects: (1) online microstructure capture and segmentation of ECCI micrographs and EBSD pattern capture; (2) offline microstructure reconstruction and robust data filling of less certain regions.

## 2.1 Online Data acquisition

The data acquisition step is performed at the SEM where instrument time may be expensive and so image processing steps at this stage are computationally cheap and kept to a minimum.

In brief, we perform the following operations online (see Figure 2):

(1) Capture a high contrast ECCI micrograph with optimised imaging conditions ('Far-field ARGUS image');
(2) Process this micrograph to segment regions of similar contrast and identify region centres;
(3) Move the EBSD detector to a position that optimises EBSD pattern capture and capture an ECCI micrograph ('Near-field ARGUS image');
(4) Register the near field and far field ECCI micrographs and capture one EBSD pattern at the centre of each region.

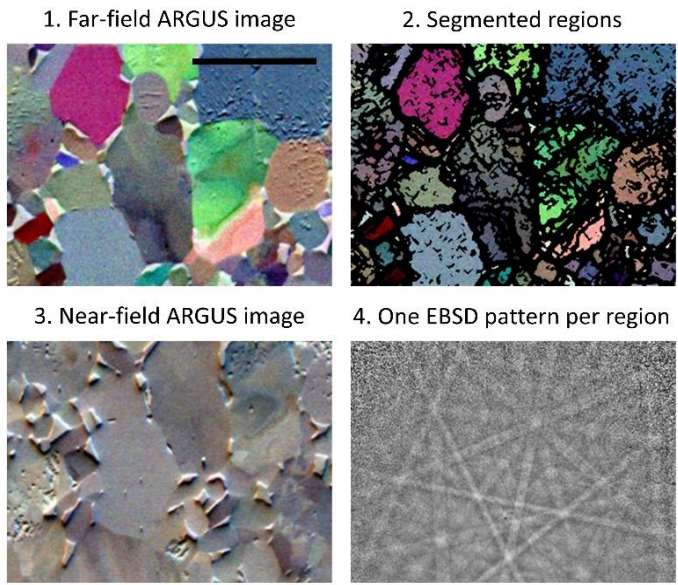

Figure 2: Flowchart showing steps during data acquisition on the microscope (the scale bar is 20 µm for this microstructure).

### (1) Orientation contrast (far field ARGUS) imaging

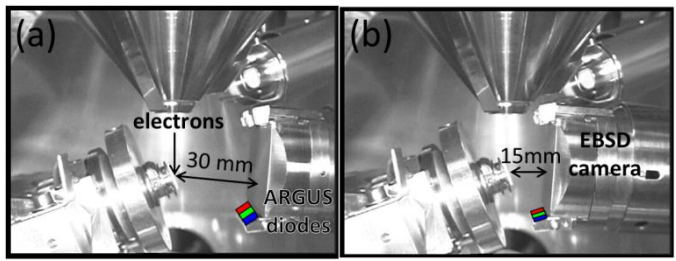

Figure 3: Infra-red images of the microscope chamber, showing ARGUS forescatter electron imaging diodes on the EBSD detector in (a) far field and (b) near-field configurations.

Orientation contrast imaging using multiple forescatter electron diodes (ARGUS diodes in the Bruker system [24]) produces a RGB false-colour micrograph using three electron detecting diodes which are located at the bottom of an EBSD detector, shown in Figure 3 (a). The signal can be tuned by optimising the detector projection geometry to produce pronounced orientation contrast and minimized topographical contrast in a highly tilted sample [3]. Both the backscattering coefficient and the anisotropy in backscattering increases with sample tilt [25], so moving the detector further back than the typical EBSD acquisition geometry, as shown in Figure 3 (b), is favourable for higher contrast in orientation contrast images. The forescatter diodes measure the total electron intensity on each 5×5 mm$^2$ diode, and signal is dominated by electron channelling-in, which produces the inelastic scattered background intensity on EBSD diffraction patterns. The origin of the ARGUS signal has been demonstrated using raw and background corrected EBSD patterns to produce analogous images [3]. Ultimately we collect an ECCI micrograph with a high signal to noise ratio using reasonable dwell times and line averaging.

**(2) Image segmentation into regions**

The far field ECCI micrograph image is processed to create a boundary map that highlights strong contrast. In practice, we perform greyscale opening and closing operations using a disk-shaped structuring element in the red, green and blue channels separately [26] and use a Gaussian smoothing filter to suppress image noise. Segmented regions of common contrast are identified by binary thresholding the 'H-concave' transformation of the boundary image [27]. Each region is assigned a unique numerical label, forming the basis of a region identifier.

Noisy and near-boundary regions are not assigned to a region in this step. The noise threshold should be set sufficiently low so that each identified region is located within one region of common phase or orientation and does not cross crystallographic boundaries, but sufficiently high so that every grain interior contains at least one labelled region. A grain can be covered by more than one region, though this increases EBSD data acquisition time. EBSD patterns are captured from region centres, which is the point with highest Euclidean distance from its region's boundaries.

Simpler image processing pipe-lines, such as thresholding and particle detection, could be used to identify regions but we did not find these to be as robust for variable boundary contrast (such as in the Co-based superalloy example shown here).

**(3) Near field ARGUS imaging**

The EBSD detector is moved inwards to the desired position for fast EBSD pattern capture (10-15 mm detector distance). A faster near field ARGUS micrograph (saved as 24 bit colour), with reduced electron channelling contrast, is captured and registered against the far field micrograph to account for distortions due to movement of the EBSD detector into the chamber. These spatial image shifts can involve deflection of the beam with respect to the sample of a few micrometres and involve affine image distortions. Registration ensures that the region centre locations determined previously still represent the same regions. Registration is performed using whole field digital image correlation, fitting displacements to an affine transform.

**(4) EBSD data acquisition**

At this stage, we now have a list of representative points that we can use to describe the orientations of our segmented regions. EBSD patterns are captured for each region and indexed. In this work, EBSD patterns are captured and saved to disk using the scripting interface in Bruker ESPRIT 2.1 [28]. If the sampled EBSD pattern is not indexed well, then another point can be selected within this region for on-line indexing. Furthermore, drift correction between capturing each EBSD patterns could be applied for more complex samples, but we note that the EBSD pattern capture happens immediately after the near field image capture and thus the impact of microscope or specimen drift is small."

## 2.2 Offline image reconstruction

At this stage, we have an incomplete 2D micrograph containing many points that are not described by a region (see Figure 2-2) because the fast segmentation algorithm assigns unique labels to unconnected 'island' regions. However, close up examination of the far field ECCI micrograph (Figure 2-1) reveals that the majority of unassigned points can be attributed to a nearby region. The fast segmented far field ARGUS micrograph typically has ~80 % of points assigned to labelled regions. The other ~20 % of points are either noisy or near- boundary points, but most can still be identified as belonging to a region. Reconstruction of the unassigned points to a region is computationally more expensive and not required for data collection, therefore it is included as a postprocessing step. After reconstruction, up to 99 % of points are assigned to labelled regions.

We perform an iterative region growing operation using selective edge dilation, where the outer edge points of each region are compared in colour to each region within the map. If they are sufficiently close in RGB vector space to the nearest and connected region they are assigned to that region. Regions are only grown if there are points which reasonably correlate to their neighbourhood (unlike a simple edge dilation which would result in morphological artefacts). A relative measure of similarity is used – the unassigned point must be more similar in colour to the candidate region than the other regions around it. This avoids the need for any hard threshold dependent on image contrast, which would need to be calibrated per image. This step is iterated until the entire map is consumed or there are no further points which fulfil the region growing criteria. Figure 4 shows a schematic of this assignment process).

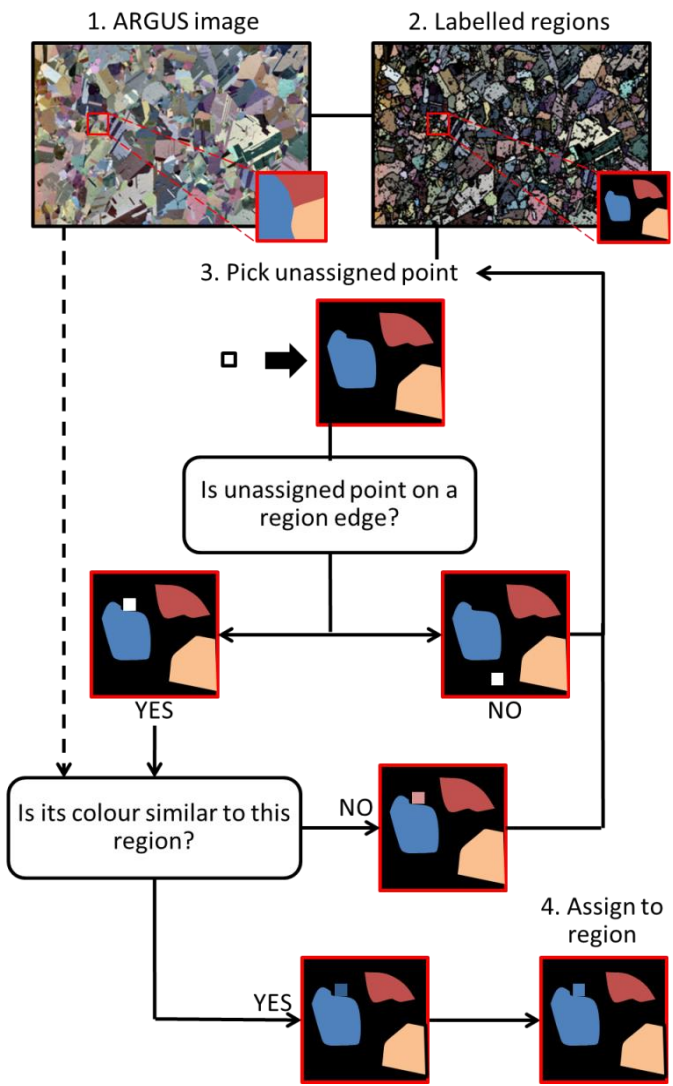

**Figure 4: Decision tree showing image reconstruction method.**

At this stage we now have a labelled region map and a series of representative region measurements for each region. A 2D EBSD map can now be reconstructed by combining the labelled region map with the representative orientation data collected for each region.

## 3 Experimental Demonstration

Experimental datasets were acquired on a Zeiss Sigma 300 field emission gun scanning electron microscope (FEG-SEM) [29], using a Bruker *e*-Flash[HD] detector fitted with ARGUS forescatter electron imaging diodes [24]. Image segmentation, offline reconstruction, and flood-filling of orientation data were performed in MATLAB [30], then visualised using the MTEX analysis toolbox [31]. EBSD patterns were acquired using the scripting interface in Bruker ESPRIT 2.1 software [28], and indexed offline with the open-source AstroEBSD algorithm [32]. The SEM was operated at a high accelerating voltage and probe current to maximise signal to noise ratio and minimise the influence of near-surface defects during ARGUS imaging and EBSD acquisition. Acquisition parameters for these two samples are given in Table 1.

For comparison, full raster maps of these microstructures were also captured using the conventional EBSD approach (i.e. measurement at each point in an X-Y grid) using the same microscope conditions.

No clean-up of the EBSD data was performed within MTEX.

|  | **Co-based superalloy** | **Ti-64** |
|---|---|---|
| **Accelerating voltage (kV)** | 30 | 20 |
| **Probe current (nA)** | ~10 | ~10 |
| **Working distance (mm)** | 14.5 | 18.0 |
| **Detector distance (mm) - Far-field ARGUS** | 30 | 30 |
| **Effective dwell time (μs) – Far-field ARGUS** | 128 | 128 |
| **Imaging time (s) – Far-field ARGUS** | 96 | 96 |
| **Detector distance (mm) - Near-field ARGUS, EBSD** | 17 | 15 |
| **Effective dwell time (μs) - Near-field ARGUS, EBSD** | 8 | 8 |
| **Imaging time (s) - Near-field ARGUS** | 8 | 8 |
| **EBSP binning – Rapid EBSD** | 8 × 8 | 5 × 5 |
| **EBSP exposure time (ms) – Rapid EBSD** | 30 | 43 |
| **EBSP binning – Conventional EBSD** | 16 × 16 | 10 × 10 |
| **EBSP exposure time (ms) – Conventional EBSD** | 7.6 | 10 |
| **ARGUS image size (Far-field and near-field)** | 750 rows × 1000 columns × (0.2 μm)$^2$ | 750 rows × 1000 columns × (0.15 μm)$^2$ |
| **Conventional EBSD map size** | 750 rows × 1000 columns × (0.2 μm)$^2$ | 750 rows × 1000 columns × (0.15 μm)$^2$ |
| **Number of Rapid EBSD regions** | 3425 | 6009 |

Table 1: Acquisition parameters for the Rapid EBSD and conventional EBSD maps.

We tested the Rapid EBSD method on two samples: a powder metallurgy processed cobalt-based superalloy containing narrow annealing twins inside larger grains, and unidirectionally rolled Ti-6Al-4V (Ti-64) with orientation gradients and a bimodal grain size distribution. Both samples were mechanically polished by standard metallographic routes to finish suitable for EBSD with low surface strain and roughness (finishing with a 0.05 μm colloidal silica slurry). These samples were chosen to test the usefulness of Rapid EBSD in correlating spatial and orientation information in microstructures, and its robustness in the presence of orientation gradients and dual phase materials.

Grain size distributions were calculated from Rapid and conventional EBSD maps using the MTEX toolbox [31]. Very small grains (grain area ≤ 15 pixels for the high spatial resolution EBSD maps) were discarded as these were noisy regions, usually located near carbides in the Co-based superalloy. In addition, a spatially binned EBSD map was produced from conventional EBSD data to compare with sparse sampling. The spatially binned maps contained a similar number of sampling points to the Rapid EBSD dataset, but in a regular grid. Single pixel grains in the binned maps were discarded, as they would not be easily distinguishable from misindexed points in an experimental dataset.

# 4 Results

## 4.1 Cobalt-based superalloy with narrow twins

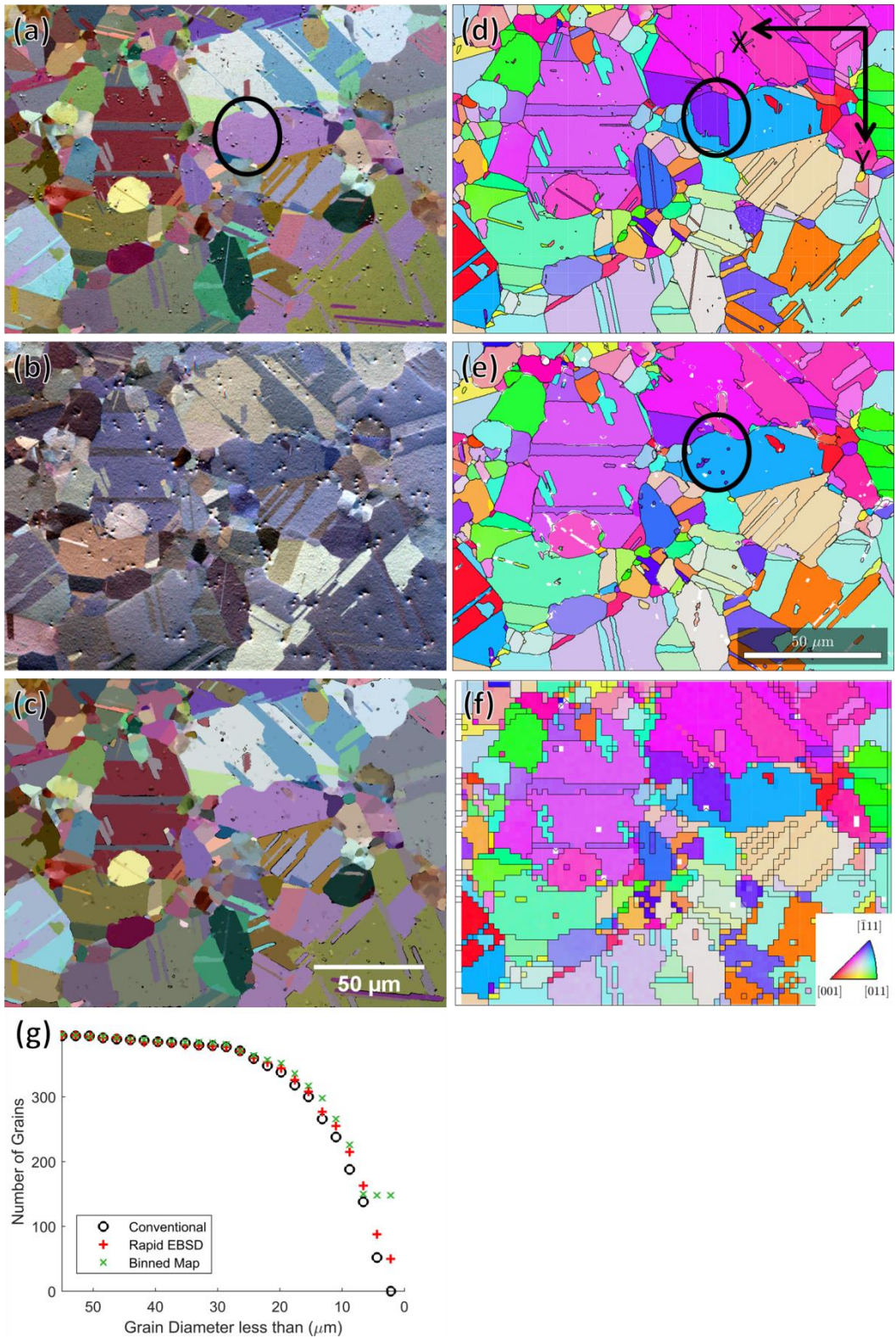

**Figure 5: Rapid EBSD dataset for cobalt-based superalloy. (a) Far-field ARGUS image; (b) Near-field ARGUS image; (c) Reconstructed far-field ARGUS regions filled in with the average ARGUS colour; (d) conventional EBSD map (IPF-X); (e) Rapid EBSD map (IPF-X); (f) spatially binned EBSD map (IPF-X); (g) grain size comparison plots using the conventional EBSD map as a reference microstructure.**

Figure 5(a) shows the far-field ARGUS image of a Co-based superalloy sample with good intergranular colour contrast from differences in the electron channelling-in signal. Figure 5(b) shows the near-field ARGUS image with comparatively supressed orientation contrast and increased topography contrast from carbides.

Figure 5 (c) shows the reconstructed far-field ARGUS image, where each of the 3425 regions is flood-filled with the average far-field ARGUS colour of the reconstructed region. 99 % of points were assigned to a region during the reconstruction step, and unassigned points are shown in black. Figure 5 (c) compares favourably with Figure 5 (a).

Figure 5 (d) shows a conventional EBSD map (inverse pole figure map along X with grain boundaries overlaid) acquired from a 1000×750 point grid with a 0.2 µm step size. Figure 5 (e) shows the equivalent Rapid EBSD map, reconstructed by assigning a single representative orientation to each reconstructed region in Figure 5 (c). The maps were cropped by 20 % to show similar fields of view.

97 % of Rapid EBSD datapoints were successfully indexed (mean angular error < 5°), compared to 99.7 % of points in the conventional map. This includes the 1% of points not assigned during the reconstruction step (black points in Figure 5 (c)), and points at twins and near grain boundaries which were assigned to regions but where the acquired pattern could not be indexed. (Bruker ESPRIT 2.1 [28] was used to index the conventional EBSD map, and AstroEBSD [32] was used to index the Rapid EBSD data.)

Figure 5 (f) shows the result from spatially binning the conventional EBSD map, increasing the step size by 10× to 2 µm. The grain shapes and boundary morphologies are poorly resolved and some narrow twins are not connected.

Figure 5 (g) shows cumulative frequency plots of the grain diameters for the conventional EBSD map, Rapid EBSD map and binned EBSD map, shown in Figure 5 (d), (e) and (f) respectively. A boundary threshold angle of 5° was chosen, and grains were similarly segmented in all maps. The conventional EBSD map in Figure 5 (d) is used as a reference for the 'true' grain structure. The grain diameter of larger grains are well-reproduced in all maps, but smaller grains less than 6 µm in diameter are systematically missed, corresponding to about 150 grains in this map.

A twin grain, circled in black in Figure 5 (d), is absent from the Rapid EBSD map in Figure 5 (e). Figure 5 (a) shows that this twin is the same colour as its parent in the far-field ARGUS image, and was therefore assigned to a single region in Figure 5 (c).

## 4.2 Ti-64 with orientation gradients and bimodal grain structure

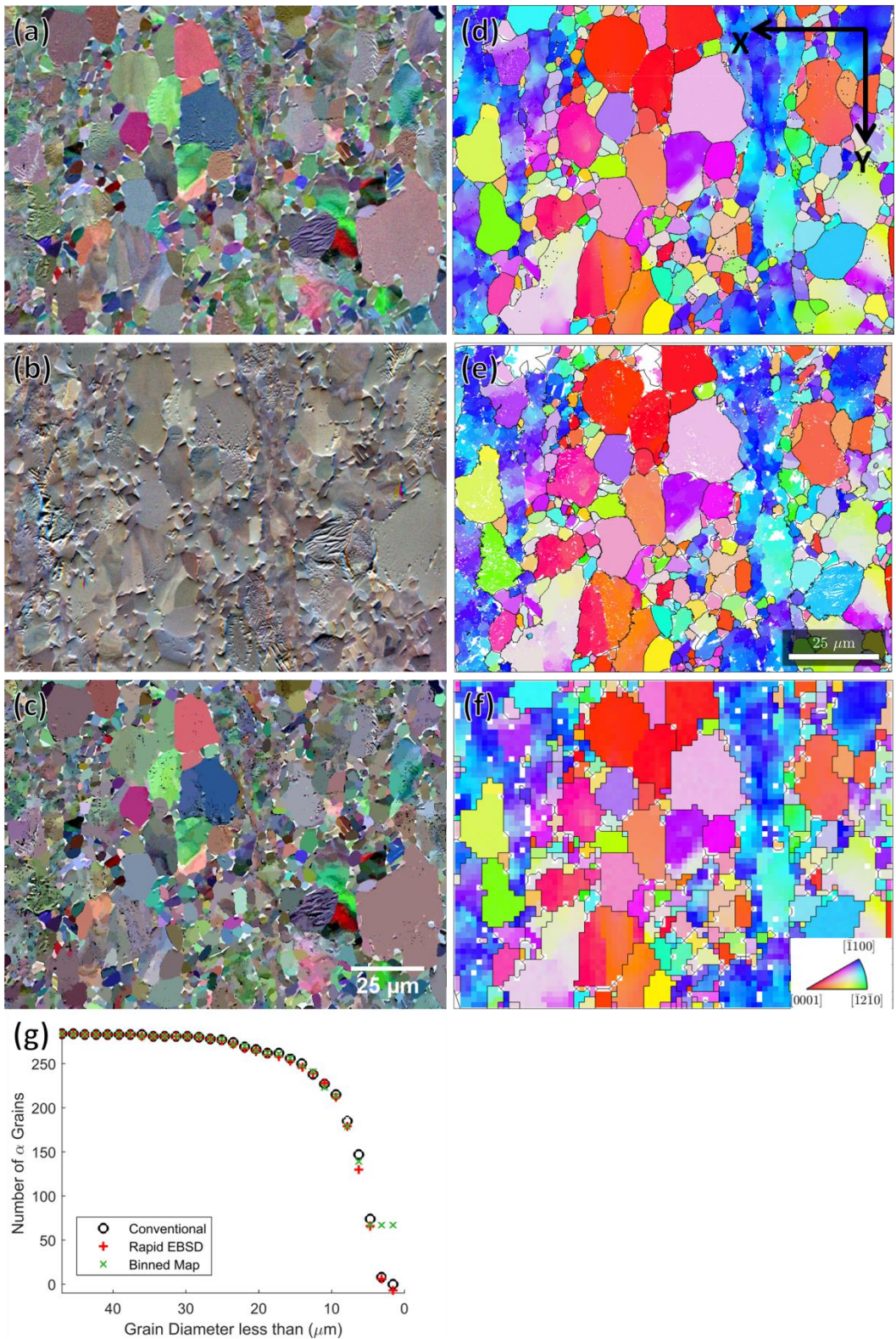

**Figure 6: Rapid EBSD dataset for Ti-6Al-4V. (a) Far-field ARGUS image; (b) Near-field ARGUS image; (c) Reconstructed far-field ARGUS regions filled in with the average ARGUS colour; (d) conventional EBSD map (IPF-Y, hexagonal α phase); (e) Rapid EBSD map (IPF-Y, hexagonal α phase); (f) spatially binned EBSD map (IPF-Y, hexagonal α phase); (g) grain size comparison plots using the conventional EBSD map as a reference microstructure.**

Figure 6 (a) shows the far-field ARGUS image of a Ti-64 sample with 97 % α phase. In addition to high intergranular colour contrast, the far-field ARGUS image also produces weaker colour variations in α grains corresponding to intragranular orientation gradients and low angle boundaries. The vanadium-rich β phase decorates triple junctions and grain boundaries. Since it has higher mean atomic number and therefore higher backscattering coefficient, it appears near-white in the ARGUS image. The near-field ARGUS image (Figure 6 (b)) shows α grains preferentially etched from mechanical polishing, leaving β protrusions.

97 % of points were assigned to one of 6009 regions in the reconstructed image shown in Figure 6 (c).

Figure 6 (d) shows a conventional EBSD map (inverse pole figure map along Y with grain boundaries overlaid) acquired from 1000×750 points with a 0.15 µm step size, and Figure 6(e) shows the equivalent Rapid EBSD map. Both EBSD maps were cropped by 20 % to show similar fields of view.

92 % of Rapid EBSD datapoints were successfully indexed, compared to 98.6 % of points in the conventional map. As 97 % of points were assigned during ARGUS image reconstruction, the gap in indexing success rate shows that the Bruker algorithm outperforms AstroEBSD, despite the longer exposure time used for AstroEBSD (Table 1).

Figure 6 (f) shows the result from spatially binning the conventional EBSD map, increasing the step size by 10× to 2 µm. Similarly to Figure 5 (f), grain shapes and boundary morphologies are poorly resolved.

Figure 6 (g) shows the cumulative frequency plots of the α-phase grain diameter for the conventional, Rapid and binned EBSD maps. The trends match the cobalt-based superalloy data (Figure 5 (g)), where grain diameters less than 5 µm are missed in the binned EBSD map, corresponding to around 70 α grains. The Rapid EBSD and conventional maps produce similar grain size distributions at all grain sizes.

Different angular thresholds were needed to produce equivalent grain boundary structures in all maps, as there were low-angle boundaries and intragranular rotation gradients in the sample. A 10° threshold was used for the conventional and Rapid EBSD maps, and 12° for the binned EBSD map. The maximum grain diameters for all maps were around 95 µm and corresponded to the same regions on the map. Using a 10° threshold for the binned EBSD map reduced the maximum grain diameter to 63 µm, because the largest grain was fragmented by low angle boundaries. This phenomenon is analogous to the increasing sensitivity of GND density measurement with increasing EBSD step size in high angular resolution EBSD [33].

## 5   Discussion

Figure 5 and Figure 6 demonstrate that Rapid EBSD can replicate conventional EBSD data and reproduce the grain shape and size distribution with better accuracy than sampling an equivalent number of data points on a coarse square grid.

The cobalt-based superalloy sample has low surface roughness, limited surface topography, high angular anisotropy in the backscatter coefficient, and negligible intragranular orientation gradients (modal kernel average misorientation of 0.3°, limited by EBSD angular resolution). This makes it a model material for image segmentation and reconstruction. The Ti-64 samples demonstrates that Rapid EBSD is robust even for more challenging materials which have significant surface topography, intragranular colour variations in the far field ARGUS image, and lower channelling contrast.

The image segmentation and reconstruction algorithms used here (edge detection and H-minima filtering, selective edge dilation, and flood-filling) were suitable for these ARGUS images, but are not necessarily optimal for all imaging modes and they could likely be refined. We have opted for fast and robust algorithms for the offline operations to render a usable solution, however any appropriate image segmentation and reconstruction algorithm can be used to inform the static sampling approach used within Rapid EBSD.

## 5.1 Benefits

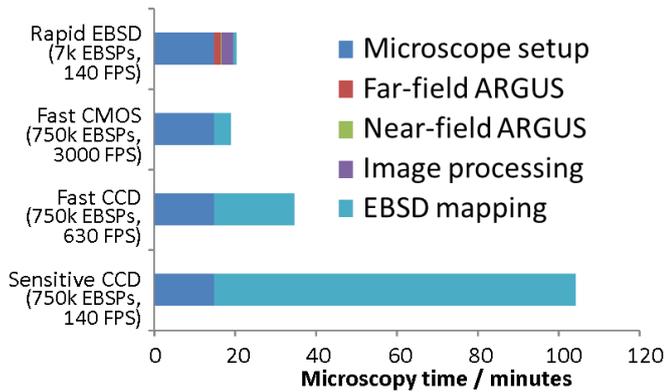

Figure 7: Comparison of microscopy time required for Rapid EBSD, and conventional mapping with a range of EBSD detector speeds between 140 and 3000 frames per second (FPS).

The main benefit of this technique is that it saves instrument time, enabling many more samples or larger fields of view to be analysed on a single instrument. The benefit is most pronounced in microstructures with heterogeneous grain structures such as those with bimodal grain size distributions. As the angular sensitivity of ECCI (ARGUS) imaging is higher than EBSD [3], the segmentation is usually conservative. Intragranular orientation gradients are approximately reproduced in many grains of the Ti-64 sample (Figure 6).

A Rapid EBSD dataset similar to the cobalt-based superalloy and Ti-64 data should take around 10 minutes to acquire: far field ARGUS image acquisition (< 2 minutes), near field ARGUS image acquisition (8 s), optimising image segmentation (around 30 s per segmentation calculation in MATLAB and 10s in C++, total < 3 minutes), and the potential to acquire EBSD patterns from the beam position list rapidly (<1 minute for a map of 6000 patterns, at 140 patterns per second). In contrast, the acquisition time for the EBSD map in Figure 5(d) was around 1.5 hours (1000×750 points at 130 patterns per second). Figure 7 shows that the total data acquisition time of Rapid EBSD using a sensitive charge-coupled device (CCD) detector (such as Bruker *e*-Flash$^{HD}$) is comparable to the mapping time of fastest currently available detectors using complementary metal-oxide-semiconductor (CMOS) technology [34] and therefore our estimation is of the potential speed up is likely conservative.

## 5.2 Limitations

Due to the qualitative nature of orientation contrast imaging, processing a single micrograph cannot ensure that all grains are segmented. In a grain imaging study comparing qualitative grain imaging with EBSD orientation maps, Day and Quested showed that with colour forescatter electron imaging (equivalent to the ARGUS imaging presented here), 3% of grain boundaries are typically missed due to similarity in the channelling conditions. Conventional low sample tilt backscatter electron and optical imaging methods miss even more boundaries [14]. An example of a missed boundary is shown in Figure 5, where a twin grain in the conventional EBSD map (black circle in Figure 5 (d)) is invisible in the far-field ARGUS image (Figure 5 (a)), and therefore missed in the Rapid EBSD map (Figure 5 (e)). The identical channelling-in conditions for these grains is likely related to the symmetry of the {111} 60° twin orientation relation. In theory this problem can be addressed through capturing multiple ECCI maps with different channelling-in conditions to improve contrast [3,35].

## 5.3 Potential applications

The sampled fraction needed for successful reconstruction of a Rapid EBSD map is typically <1%, compared to 6 and 10% for existing sparse sampling methods previously discussed. Rapid EBSD may therefore afford significant advantages for experiments where data acquisition time is a limiting factor (e.g. large area map or focussed ion beam serial sectioning 3D-EBSD) or where sample or beam drift produces significant artefacts in the data (e.g. at high magnifications in transmission Kikuchi diffraction).

Rapid EBSD can also be advantageous for analysis of beam sensitive materials, as the electron dose required for good ARGUS imaging is lower than for conventional EBSD analysis. Although for this demonstration we have selected to use a 128 µs dwell time for far-field ARGUS imaging (detailed in Table 1) to minimise signal-to-noise ratio, we have also had significant success segmenting images acquired with 32 µs dwell time (a 4× reduction of electron dose and increase in imaging speed).

The underpinning idea of fusing segmented ECCI micrographs with EBSD measurements is not limited to EBSD data and we look forward to exploring combining this smart sampling approach with energy dispersive X-ray spectroscopy, where a backscatter electron image optimised to show compositional differences can be segmented into regions.

# 6 Conclusions

- We have developed an image segmentation and reconstruction algorithm to enable EBSD mapping from a sparse dataset with around 1 % sampled fraction.
- We have demonstrated that Rapid EBSD method can successfully reproduce EBSD orientation maps in a cobalt-based superalloy and titanium alloy.
- In time-limited experiments, Rapid EBSD enables grain shapes to be better resolved and grain size distributions to be measured with higher accuracy than regular grid-based mapping.

# 7 Data Statement

Data from this manuscript will be released into a Zenodo repository when this article is accepted.

# 8 Author contributions

VT and TBB developed the algorithm and explored the case studies shown here. Development of the original idea was proposed by AK in consultation with VT, TBB and DD. The manuscript was drafted by VT. All authors have contributed to the final manuscript.

# 9 Acknowledgements

We would like to thank Ken Mingard (NPL) as well as Thomas Schwager and Daniel Goran (Bruker) for helpful discussions in development of the Rapid EBSD. VT and TBB acknowledge funding from Imperial College London and HEIF through the Proof of Concept scheme. DD, VT and TBB acknowledge funding from EPSRC EP/K034332/1 (HexMat). DD and AK acknowledge funding from EPSRC EP/L025213/1 (DARE). TBB acknowledges funding of his Research Fellowship from the Royal Academy of Engineering.

# 10 References

[1] A.J. Wilkinson, T.B. Britton, Strains, planes, and EBSD in materials science, Mater. Today. 15 (2012) 366–376. doi:10.1016/S1369-7021(12)70163-3.

[2] D. Dingley, Progressive steps in the development of electron backscatter diffraction and orientation imaging microscopy, J. Microsc. 213 (2004) 214–224. doi:10.1111/j.0022-2720.2004.01321.x.

[3] B. Britton, D. Goran, V. Tong, Space rocks and optimising scanning electron channelling contrast, (2018).

[4] S.I. Wright, M.M. Nowell, R. De Kloe, P. Camus, T. Rampton, Electron imaging with an EBSD detector, Ultramicroscopy. 148 (2015) 132–145. doi:10.1016/j.ultramic.2014.10.002.

[5] ASTM International, ASTM E112 Standard Test Methods for Determining Average Grain Size, 13 (1996) 1–28. doi:10.1520/E0112-13.1.4.

[6]   R. Ohbuchi, M. Aono, Masaki Aono and Ryutarou Ohbuchi, Quasi-Monte Carlo Rendering with Adaptive Sampling, 1996.

[7]   T. Tuma, J. Lygeros, V. Kartik, A. Sebastian, A. Pantazi, High-speed multiresolution scanning probe microscopy based on Lissajous scan trajectories, Nanotechnology. 23 (2012). doi:10.1088/0957-4484/23/18/185501.

[8]   Z. Hussain, A. Muhammad, Sample size reduction in groundwater surveys via sparse data assimilation, 2013 10th IEEE Int. Conf. Networking, Sens. Control. ICNSC 2013. (2013) 176–182. doi:10.1109/ICNSC.2013.6548732.

[9]   A. Stevens, L. Luzi, H. Yang, L. Kovarik, B.L. Mehdi, A. Liyu, M.E. Gehm, N.D. Browning, A sub-sampled approach to extremely low-dose STEM, Appl. Phys. Lett. 112 (2018). doi:10.1063/1.5016192.

[10]  Z. Saghi, M. Benning, R. Leary, M. Macias-Montero, A. Borras, P.A. Midgley, Reduced-dose and high-speed acquisition strategies for multi-dimensional electron microscopy, Adv. Struct. Chem. Imaging. 1 (2015) 7. doi:10.1186/s40679-015-0007-5.

[11]  T.E. Merryman, J. Kovačević, An adaptive multirate algorithm for acquisition of fluorescence microscopy data sets, IEEE Trans. Image Process. 14 (2005) 1246–1253. doi:10.1109/TIP.2005.855861.

[12]  Y. Zhang, G.M.D. Godaliyadda, N. Ferrier, E.B. Gulsoy, C.A. Bouman, C. Phatak, SLADS-Net: Supervised Learning Approach for Dynamic Sampling using Deep Neural Networks, (2018).

[13]  Y. Zhang, G.M.D. Godaliyadda, N. Ferrier, E.B. Gulsoy, C.A. Bouman, C. Phatak, Reduced electron exposure for energy-dispersive spectroscopy using dynamic sampling, Ultramicroscopy. 184 (2018) 90–97. doi:10.1016/j.ultramic.2017.10.015.

[14]  A.P. Day, T.E. Quested, A comparison of grain imaging and measurement using horizontal orientation and colour orientation contrast imaging, electron backscatter pattern and optical methods, J. Microsc. 195 (1999) 186–196. doi:10.1046/j.1365-2818.1999.00571.x.

[15]  T. a Bennett, P.N. Kalu, A.D. Rollett, Strain-induced selective growth in 1.5% temper-rolled Fe;1%Si., Microsc. Microanal. 17 (2011) 362–7. doi:10.1017/S1431927611000377.

[16]  H. Park, D.Y. Kim, N.M. Hwang, Y.C. Joo, C.H. Han, J.K. Kim, Microstructural evidence of abnormal grain growth by solid-state wetting in Fe-3%Si steel, J. Appl. Phys. 95 (2004) 5515–5521. doi:10.1063/1.1712012.

[17]  W.T. Roberts, PRE } ~ ERRED ORIENTATION IN WROUGHT Dillamore and Roberts :, 10 (1965).

[18]  J. Dennis, P.S. Bate, F.J. Humphreys, Abnormal grain growth in Al–3.5Cu, Acta Mater. 57 (2009) 4539–4547. doi:10.1016/j.actamat.2009.06.018.

[19]  Y.S. Sato, H. Watanabe, H. Kokawa, Grain growth phenomena in friction stir welded 1100 Al during post-weld heat treatment, Sci. Technol. Weld. Join. 12 (2007) 318–323. doi:10.1179/174329307X197575.

[20]  K.A.A. Hassan, A.F. Norman, D.A. Price, P.B. Prangnell, Stability of nugget zone grain structures in high strength Al-alloy friction stir welds during solution treatment, Acta Mater. 51 (2003) 1923–1936. doi:10.1016/S1359-6454(02)00598-0.

[21]  T.W. Na, H.K. Park, C.S. Park, J.T. Park, N.M. Hwang, Misorientation angle analysis near the growth front of abnormally growing grains in 5052 aluminum alloy, Acta Mater. 115 (2016) 224–229. doi:10.1016/j.actamat.2016.06.007.

[22]  D.L. Gray, Recovery and recrystallization of zirconium and its alloys, (1961).

[23]  V.S. Tong, T. Ben Britton, Formation of very large 'blocky alpha' grains in Zircaloy-4, Acta Mater. 129 (2017) 510–520. doi:10.1016/j.actamat.2017.03.002.

[24]  QUANTAX EBSD - Overview, Analysis System - QUANTAX EBSD | Bruker, (n.d.).

[25]  L. Reimer, Scanning electron microscopy : physics of image formation and microanalysis, n.d.

[26] P. Soille, Opening and Closing, in: Morphol. Image Anal., Springer Berlin Heidelberg, Berlin, Heidelberg, 1999: pp. 89–127. doi:10.1007/978-3-662-03939-7_4.

[27] P. Soille, Segmentation, in: Morphol. Image Anal., Springer Berlin Heidelberg, Berlin, Heidelberg, 1999: pp. 229–254. doi:10.1007/978-3-662-03939-7_9.

[28] Bruker, Esprit 2.1, (n.d.).

[29] ZEISS SIGMA field emission scanning electron microscope, (n.d.).

[30] https://uk.mathworks.com/products/matlab.html, (n.d.).

[31] R. Hielscher, MTEX 5.0.3, (2018).

[32] T.B. Britton, V. Tong, J. Hickey, A. Foden, A. Wilkinson, AstroEBSD: exploring new space in pattern indexing with methods launched from an astronomical approach, (2018).

[33] J. Jiang, T.B.B. Britton, A.J.J. Wilkinson, Measurement of geometrically necessary dislocation density with high resolution electron backscatter diffraction: Effects of detector binning and step size, Ultramicroscopy. 125 (2012) 1–9. doi:10.1016/j.ultramic.2012.11.003.

[34] J. Goulden, P. Trimby, A. Bewick, The Benefits and Applications of a CMOS-based EBSD Detector, Microsc. Microanal. 24 (2018) 1128–1129. doi:10.1017/S1431927618006128.

[35] C. Lafond, T. Douillard, S. Cazottes, P. Steyer, C. Langlois, Electron CHanneling ORientation Determination (eCHORD): An original approach to crystalline orientation mapping, Ultramicroscopy. 186 (2018) 146–149. doi:10.1016/J.ULTRAMIC.2017.12.019.